\documentclass{myphbauth}       
\usepackage{graphicx}
\usepackage{graphics}
\usepackage{epsfig}
\usepackage{amssymb}

\begin{document}

\begin{frontmatter}

\title{Bond-Dilution Effects on Two-Dimensional Spin-Gapped
Heisenberg Antiferromagnets}

\author[a]{Chitoshi Yasuda}\ead{cyasuda@issp.u-tokyo.ac.jp}, 
\author[a,b]{Synge Todo}, 
\author[a]{Munehisa Matsumoto}, and \author[a]{Hajime Takayama} 

\address[a]{Institute for Solid State Physics, University of Tokyo, Kashiwa 277-8581, Japan}
\address[b]{Theoretische Physik, Eidgen\"ossische Technische Hochschule,
   CH-8093 Z\"urich, Switzerland}

\begin{abstract}
 Bond-dilution effects on spin-1/2 spin-gapped Heisenberg
 antiferromagnets of coupled alternating chains on a square lattice
 are investigated by means of the quantum Monte Carlo method. It is
 found that, in contrast with the site-diluted system having an
 infinitesimal critical concentration, the bond-diluted system has a
 finite critical concentration of diluted bonds, $x_{\rm c}$, above
 which the system is in an antiferromagnetic (AF) long-range ordered
 phase. In the disordered phase below $x_{\rm c}$, plausibly in the
 concentration region significantly less than $x_{\rm c}$, the system
 has a spin gap due to singlet pairs of induced magnetic moments
 reformed by the AF interactions through the two-dimensional shortest
 paths. 
\end{abstract}


\end{frontmatter}

\section{Introduction}

Impurity effects on quantum Heisenberg antiferromagnets with a spin
gap due to the topological structure of the system have attracted much
interest in relation to the impurity-induced antiferromagnetic (AF)
long-range order (LRO) observed
experimentally~\cite{hase,regnault,azuma,uchiyama}. There are two types
of disorders, i.e., site dilution and bond dilution. Extensive
theoretical and numerical works have investigated properties of
diluted Heisenberg antiferromagnets
in the one-dimensional (1D)~\cite{miyashita,eggert} and
two-dimensional (2D) systems. In the latter, however, the interchain
interactions have been treated by the mean-field-type
approximations~\cite{fukuyama}, so that the difference of these two
types of disorders cannot be naturally introduced. In the present paper
we have investigated bond-dilution effects on the spin-1/2 spin-gapped
AF Heisenberg system by treating the interchain interactions on the
equal footing as the intrachain interactions by means of the quantum
Monte Carlo (QMC) method. The result obtained clearly reveals the
different aspects between magnetic properties of the bond- and
site-diluted systems.

The site-dilution effects have been already extensively
investigated~\cite{imada,wessel,yasuda}. If spins are randomly removed
from the lattice, magnetic moments are induced at sites neighboring the removed
sites. We have called the domain of induced magnetic moments the
`effective spin'. They situate centered at sites next to each removed
spin, and their extent is given by the correlation lengths, $\xi_{\rm
p}^{x,y}$, of the non-diluted system. The effective exchange coupling
between two effective spins centered at sites $m$ and $n$ is given by
$\tilde{J}_{mn} \propto (-1)^{|r_m-r_n+1|}{\rm exp}[-l/(\xi_{\rm p}^{x}\xi_{\rm
p}^{y})^{1/2}]$, where $l=|r_m-r_n|$ is the distance between the
effective spins~\cite{nagaosa,iino}. There exist strong correlations between
the effective spins retaining the staggeredness with respect to the original
lattice~\cite{yasuda}. Therefore, the 2D site-diluted system has an AF
LRO induced with an {\it infinitesimal} concentration of dilution at
zero temperature. 

On the other hand, if a certain bond is removed from
the lattice, the effective spins, peaks of which are located at both
ends of the diluted bond, are induced. In this case, besides the same
interaction $\tilde{J}_{mn}$ between spins next to different removed
bonds as in the site-diluted case, there exist AF interactions $J_{\rm
AF}$ between two spins at both ends of a removed strong bond through the
2D shortest paths. The latter is estimated of the order of $J'^2$, where
$J'$ is the strength of the interchain interaction and the strength of
the stronger intrachain interaction is put unity. In the small
concentration region with $x \ll 1/\xi_{\rm p}^{x}\xi_{\rm p}^{y}$,
therefore it is expected that the singlet pair is reformed by $J_{\rm
AF}$. As $x$ is increased, $\tilde{J}_{mn}$ becomes dominant to $J_{\rm
AF}$. As a result, the AF LRO is expected to occur above a {\it finite}
critical concentration $x_{\rm c}$ in the bond-diluted case. This
scenario is clearly confirmed by the present work.

\section{Model and method}

We investigate the bond-diluted quantum AF Heisenberg model of coupled
alternating chains on a square lattice described by 
the Hamiltonian
\begin{eqnarray}
   \label{ham}
   H &=&\sum_{i,j}\epsilon_{(2i,j)(2i+1,j)}
             {\bf S}_{2i,j}\cdot{\bf S}_{2i+1,j} \nonumber \\
    &+&\alpha\sum_{i,j}\epsilon_{(2i+1,j)(2i+2,j)}
             {\bf S}_{2i+1,j}\cdot{\bf S}_{2i+2,j} \\
    &+&J'\sum_{i,j}\epsilon_{(i,j)(i,j+1)}
             {\bf S}_{i,j}\cdot{\bf S}_{i,j+1} \ , \nonumber
\end{eqnarray}
where 1 and $\alpha$ ($>0$) are the AF intrachain alternating coupling
constants, $J'$ ($>0$) the AF interchain coupling constant, and ${\bf
S}_{i,j}$ the quantum spin operator with $S=1/2$ at site
($i,j$). Randomly quenched bond occupation factors
\{$\epsilon_{(i,j)(k,l)}$\} independently take either 1 or 0 with
probability $1-x$ and $x$, respectively, where $x$ is the concentratin
of diluted bonds. The pure system described by Eq. (\ref{ham}) with
$\epsilon_{(i,j)(k,l)}=1$ for all bonds is in a spin-gapped or an AF
LRO phase depending on strengths of $\alpha$ and $J'$ at zero
temperature~\cite{matsumoto}. In the $\alpha=0.5$ system which we
examine in the present work there exists the spin-gapped phase below
$J'_{\rm c}\simeq 0.55$.

The QMC simulations with the continuous-imaginary-time loop algorithm~\cite{evertz,beard,todo} are carried out
on $L\times L$ square lattices with the periodic boundary
conditions. For each sample with a bond-diluted configuration, $10^3
\sim 10^4$ Monte Carlo steps (MCS) are spent for measurement after
$500 \sim 10^3$ MCS for thermalization. Sample average for dilution is
taken over $10 \sim 10^3$ samples.

\section{Numerical results}

\begin{figure}[t]
\begin{center}\leavevmode
 \includegraphics[width=0.9\linewidth]{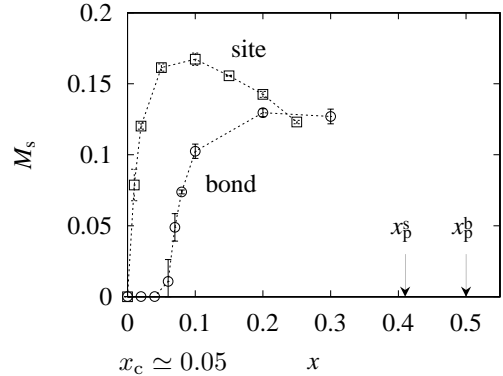}
\end{center}
 \vspace*{-0.5cm}
 \hspace*{2cm}{$x_{\rm c} \simeq 0.05$}
\caption{Concentration dependences of the staggered magnetization
	    in the bond- and site-diluted [10] systems with
	    $\alpha=J'=0.5$. There exists the critical concentration of
	    $x_{\rm c} \simeq 0.05$ in the bond-diluted system. The
	    percolation thresholds on the bond and site processes are
	    denoted by $x_{\rm p}^{\rm b}$ and $x_{\rm p}^{\rm s}$,
	    respectively. All the lines are guides to eyes.}
\end{figure}

The $x$ dependences of the staggered magnetization $M_{\rm s}$ in the bond- and
site-diluted~\cite{yasuda} systems with $\alpha=J'=0.5$ are shown in
Fig. 1. If the pure system with the spin-gapped state is diluted, the AF
LRO is induced at a certain concentration in both site- and bond-diluted
cases. However, whereas the site-diluted system is driven into the AF LRO phase
by an infinitesimal concentration, in the bond-diluted system there
exists a critical concentration of $x_{\rm c} \simeq 0.05$ even for
$\alpha=J'=0.5$, which is the system near the critical point $J'_{\rm c}
\simeq 0.55$ for $\alpha=0.5$ and $x=0$. As the value of $J'$
becomes smaller, the region of the disordered phase, i.e., the value of
$x_{\rm c}$ increases due to the reduction of $\xi_{\rm p}^{x,y}$, e.g.,
$x_{\rm c} \simeq 0.1$ for $J'=0.3$.

\begin{figure}[t]
 \begin{center}\leavevmode
  \includegraphics[width=0.9\linewidth]{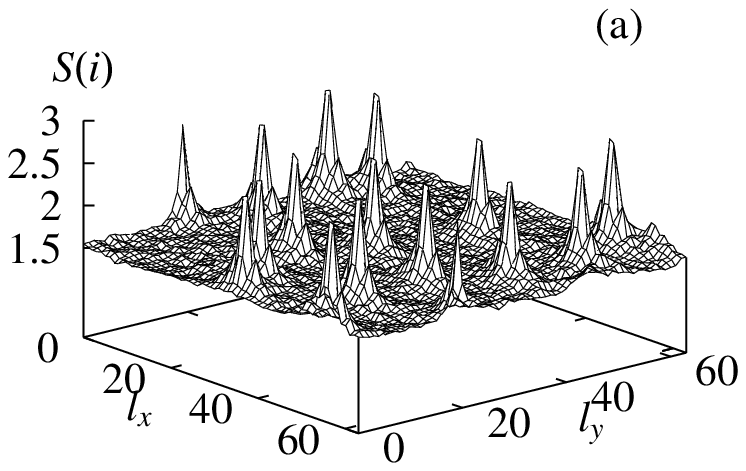}
 \end{center}
 \vspace*{-1cm}
 \begin{center}
  \includegraphics[width=0.9\linewidth]{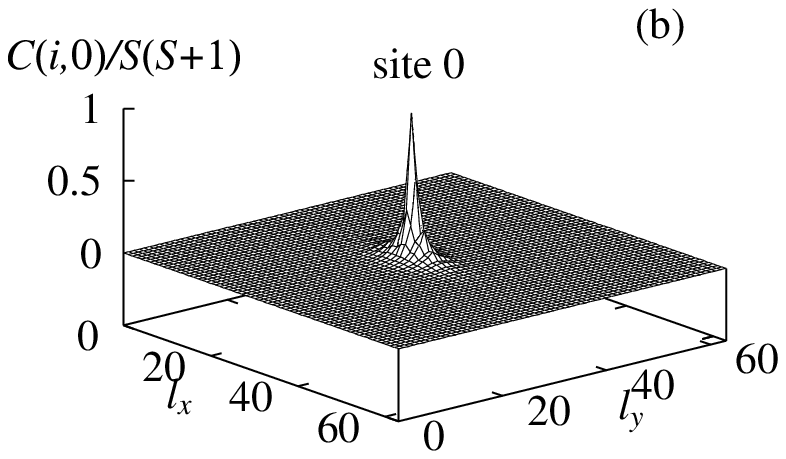}
 \end{center}
\vspace*{-0.5cm}
\caption{Real-space distribution of (a) the local static staggered
	    structure factor and (b) the staggered correlation function
	    in a fixed configuration with 15 diluted strong bonds for
	    $\alpha=0.5$ and $J'=0.3$ at $T=0.001$.}
\end{figure}

For the coupled alternating chains considered in the present work there
exist three types of bonds, whose strengths are 1, $\alpha$, and $J'$,
respectively. If a weaker bond with $\alpha$ or $J'$ is removed, the singlet
pairs on the strong bonds are not disturbed at all and so magnetic
moments are not induced. However, if a strong bond is removed, magnetic
moments are induced at both ends of the diluted bond. In Fig. 2(a) the
real-space distribution of the local static staggered structure factor
defined by $S(i)\equiv \sum_{j}(-1)^{|r_{i}-r_{j}|}\langle
S_{i}^{z}S_{j}^{z}\rangle$ is shown for $\alpha=0.5$ and $J'=0.3$ on a
64 $\times$ 64 lattice from which 15 strong bonds are removed. We can
clearly see that magnetic moments are always induced in pairs at both
ends of the removed bonds. This is a different situation from the site
dilution for which isolated magnetic moments are induced next to each
diluted site~\cite{yasuda}. Furthermore, as shown in Fig. 2(b), the
real-space distribution of the staggered correlation function,
$C(i,0)=(-1)^{|r_{i}-r_{0}|}\langle S_{i}^{z}S_{0}^{z} \rangle$,
exhibits no AF LRO, where 0 (=(29,31)) is the peak site of an effective
spin. (The reader may compare it with the AF LRO in the site-diluted
case in ref.~\cite{yasuda}). In such a small concentration region with
$x \ll 1/\xi_{\rm p}^{x}\xi_{\rm p}^{y}$, where $\xi_{\rm
p}^{x}=3.452(1)$, $\xi_{\rm p}^{y}=1.8202(3)$ for $\alpha=0.5$ and
$J'=0.3$, the spins correlate only with the neighboring spins. In the
concentration region of $x \sim 1/\xi_{\rm p}^{x}\xi_{\rm p}^{y}$, on
the other hand, the effective spins practically overlap each other and
the AF LRO is induced.

\begin{figure}[t]
\begin{center}\leavevmode
 \includegraphics[width=0.9\linewidth]{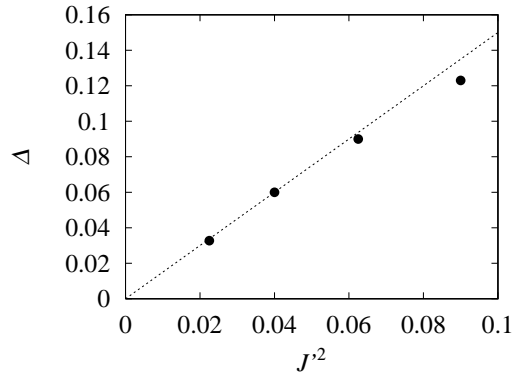}
\end{center}
\caption{Spin gap proportional to $J'^2$ in the
	    extremely-low-concentration region for $\alpha=0.5$. A
	    line with the slope of 1.5 is the guide to eyes.}
\end{figure}

Next we investigate whether or not there is the spin-gapped phase in the
disordered phase by means of the moment method~\cite{todo,cooper}. We
have found that in the extremely-low-concentration region there exists a spin
gap, $\Delta$, whose magnitude is independent of the value of $x$ and
proportional to $J'^2$ as shown in Fig. 3. This relation tells that, as
mentioned in Introduction, two edge spins induced at both ends of a
diluted bond reform a singlet pair due to the effective interaction
$J_{\rm AF}$ ($|J_{\rm AF}| \sim J'^2$) through the 2D shortest paths. We
note here that since induced magnetic moments are located at sites of
the different sublattices, the interaction is always AF. As $J'^2$ is
increased, $\Delta$ deviates from the proportionality to $J'^2$. This
indicates that the localized singlet picture does not hold well for
larger values of $J'$.

\begin{figure}[t]
\begin{center}\leavevmode
 \includegraphics[width=0.9\linewidth]{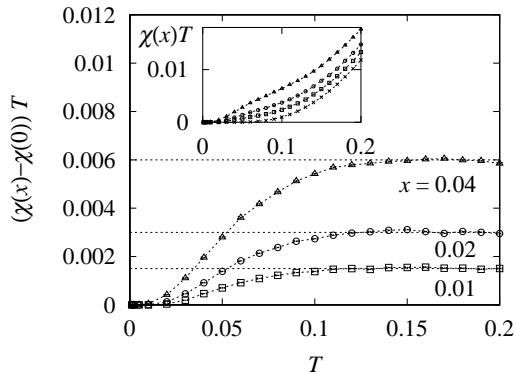}
\end{center}
\caption{Temperature dependence of the effective Curie constant
	    $(\chi(x)-\chi(0))T$ for $\alpha=0.5$ and $J'=0.3$. In the
	    inset $\chi(x)T$ are shown at $x=0.04$, 0.02, 0.01, and 0
	    from top. All the lines are guides to eyes.}
\end{figure}

In the inset of Fig. 4 the $T$ dependence of the effective Curie
constant $\chi(x) T$ for $\alpha=0.5$ and $J'=0.3$ is shown at $x=0.04$,
0.02, 0.01, and 0 from top. The uniform susceptibility $\chi(0)$
exponentially decreases at low temperatures due to the spin gap with the
value of 0.31415(2). In the main frame of Fig. 4, the differences
$(\chi(x)-\chi(0))T$ are shown. The plateau with the Curie constant
proportional to $x$ is clearly seen in these plots and
$(\chi(x)-\chi(0))T$ falls off exponentially at lower temperature than
$J'^2 = 0.09$. This confirms again the existence of the spin gap due to
$J_{\rm AF}$ at relatively small $x$.

\section{Summary and discussion}

When bonds are randomly removed in the spin-gapped phase,
the AF effective interaction $J_{\rm AF}$ between two spins at both ends
of a removed strong bond is induced in addition to the
effective interaction $\tilde{J}_{mn}$ between spins on ends of
different removed bonds. Therefore, in such a low-concentration region
that the strength of $J_{\rm AF}$ is larger than $|\tilde{J}_{mn}|$ there
exists the spin-gapped phase dominated by $J_{\rm AF}$.

The ground state of the diluted dimerized 1D system with $J'=0$ and $x>0$ is in
the quantum Griffiths phase, which is characterized by
the finite correlation length and the gapless excitation. In the present
2D system with $J'>0$ our numerical results strongly suggest a gapless
phase but without AF LRO just below $x_{\rm c}$. The problem will be
discussed elsewhere.

\begin{ack}
Most of numerical calculations in the present work have been performed
on the SGI 2800 at Institute for Solid State Physics, University of
Tokyo. The program is based on 'Looper version 2' developed by S.T. and
K. Kato and 'PARAPACK version 2' by S.T. The present work is supported
by the ``Research for the Future Program'' (JSPS-RFTF97P01103) of Japan
Society for the Promotion of Science.
\end{ack}

\end{document}